\documentclass[manuscript,screen]{acmart}
\usepackage[utf8]{inputenc}
\usepackage{textcomp}
\usepackage{newunicodechar}
\newunicodechar{−}{\textminus}
\AtBeginDocument{%
  }

\setcopyright{acmlicensed}
\copyrightyear{2018}
\acmYear{2018}
\acmDOI{XXXXXXX.XXXXXXX}
\acmConference[TEI '25]{Make sure to enter the correct
  conference title from your rights confirmation email}{June 03--05,
  2018}{Chicago, Illinois}
\acmISBN{978-1-4503-XXXX-X/2018/06}

\begin{document}

\title{Augmented Assembly: Object Recognition and Hand Tracking for\\ Adaptive Assembly Instructions in Augmented Reality}
\author{Alexander Htet Kyaw}
\affiliation{%
  \institution{Massachusetts Institute of Technology}
  \city{Cambridge}
  \state{MA}
  \country{United States}}
\email{alexkyaws@mit.edu}

\author{Haotian Ma}
\affiliation{%
  \institution{Cornell University}
  \city{Cornell}
  \state{NY}
  \country{United States}}
\email{hm443@mit.edu}

\author{Sasa Zivkovic}
\affiliation{%
  \institution{Cornell University}
  \city{Ithaca}
  \state{NY}
  \country{United States}}
\email{sz382@cornell.edu}

\author{Jenny Sabin}
\affiliation{%
  \institution{Cornell University}
  \city{Ithaca}
  \state{NY}
  \country{United States}}
\email{jes557@cornell.edu}

\renewcommand{\shortauthors}{Kyaw, Ma, Zivkovic, and Sabin}


\begin{abstract}
Recent advances in Augmented reality (AR) have enabled interactive systems that assist users in physical assembly tasks. In this paper, we present an AR-assisted assembly workflow that leverages object recognition and hand tracking to (1) identify custom components, (2) display step-by-step instructions, (3) detect assembly deviations, and (4) dynamically update the instruction based on users’ hands-on interactions with physical parts. Using object recognition, the system detects and localizes components in real time to create a digital twin of the workspace. For each assembly step, it overlays bounding boxes in AR to indicate both the current position and the target placement of relevant components, while hand-tracking data verifies whether the user interacts with the correct part. Rather than enforcing a fixed sequence, the system highlights potential assembly errors and interprets users deviations as opportunities for iteration and creative exploration. A case study with LEGO-blocks and custom 3D-printed-components demonstrates how the system links digital instructions to physical assembly, eliminating the need for manual searching, sorting, or labeling of parts.

\end{abstract}


\begin{teaserfigure}
  \includegraphics[width=\textwidth]{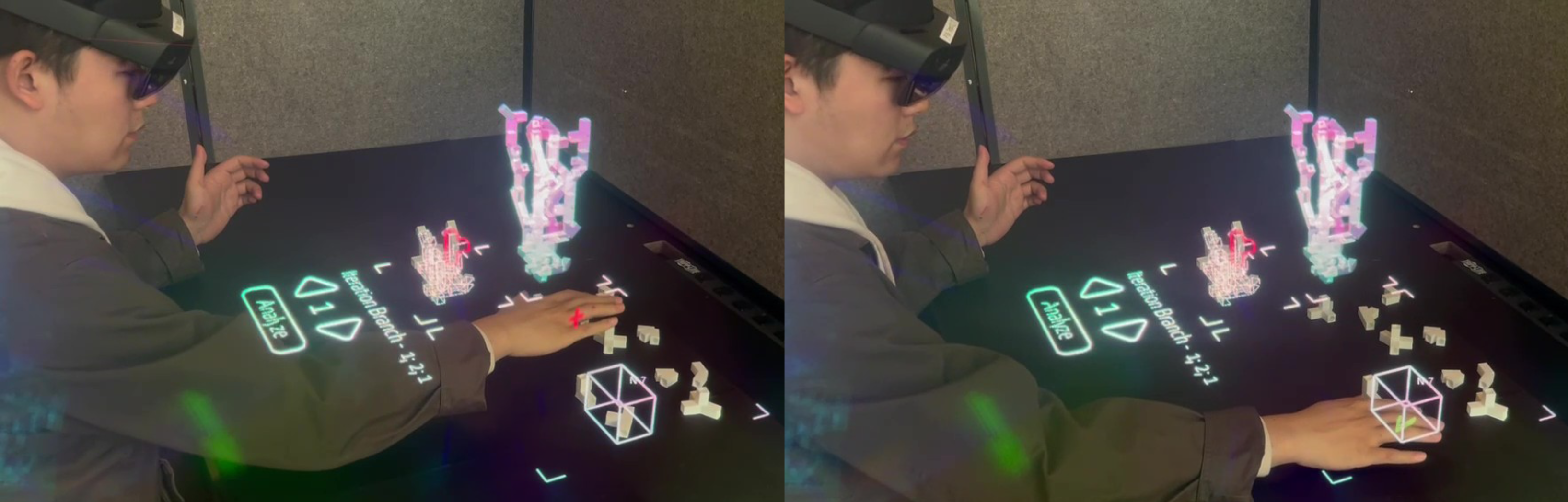}
  \caption{Object recognition and hand tracking in augmented reality provide visual feedback during the physical assembly process. For each step, a bounding box highlights the relevant component, a red outline indicates its target placement, and the full assembly model is displayed in front. In the left image, the system displays a red cross when the user’s hand hovers over a component outside the current assembly step, while in the right image, a green check mark confirms interaction within the assembly step.}
  \Description{teaser}
  \label{fig:teaser}
\end{teaserfigure}


\maketitle

\section{Introduction}

Assembling physical objects is an embodied and tangible process that integrates perception, coordination, and spatial reasoning through direct interaction with materials and space. While digital fabrication and design tools have advanced rapidly, the act of physical assembly still relies heavily on static instructions, paper manuals, videos, or digital models, that separate guidance from action \cite{lee-cultura_embodied_2020, yang_comparing_2020}. This disconnect often leads to errors, inefficiency, and fragmented attention, especially when users must interpret 2D information for 3D tasks \cite{henderson_augmented_2011}. 

Emerging research in augmented and mixed reality has begun to address these challenges by overlaying digital guidance directly within the user’s environment  \cite{wang_comprehensive_2022, han_immersive_2025, kang_prototyping_2018, hattab_interactive_2019} . Yet, many existing systems remain limited by fixed component tracking or pre-registered models, offering little adaptability to real-world variations (Wei et al., 2023) \cite{fang_research_2023-1, lee_design_2023}. At the same time, advances in computer vision and machine learning present new opportunities for making assembly processes more responsive, interactive, and context-aware \cite{garcia_de_paiva_pinheiro_integrating_2025}

In this work, we explore how real-time spatial understanding of assembly components, combined with users’ hand interactions, can enhance embodied and situated making \cite{ma_ai_2023, kyaw_ai_2025}. Our system integrates object recognition and hand tracking within an augmented reality workflow to identify physical assembly components, monitor real-world user interactions, and dynamically update the digital assembly instructions accordingly (Figure. \ref{fig:teaser}). The system not only visualizes assembly deviations but also adapts the assembly instructions to accommodate user improvisation and creative intent in the physical act of making. By linking digital feedback to physical action, the system supports adaptive guidance that responds to user behavior instead of enforcing a fixed sequence. The system demonstrates:
\begin{itemize}
    \item An AR assembly workflow that integrates object recognition and hand tracking to create a digital twin and provide step-by-step instructions aligned with the user’s physical interactions and real-world assembly components.
    \item A feedback mechanism that highlights errors and detects deviations to dynamically adapt instructions based on the physical assembly state, and allows users to explore assembly alternatives guided by structural analysis.
    \item A case study showing how the system removes the need for manual searching, sorting, or labeling of parts during physical assembly, using LEGO blocks and custom 3D-printed components as examples.
\end{itemize}

\section{Related Work}
AR has been used to overlay digital instructions directly onto the physical workspace to support embodied assembly and fabrication task \cite{chen_augmenting_2024, liu_instrumentar_2023, funk_working_2017}. Prior work demonstrates that AR can enhance task comprehension by visualizing part placement, sequencing, and spatial orientation to reduce cognitive load during assembly \cite{yuan_augmented_2024, alves_comparing_2022}. Recent studies on using AR for task guidance emphasize the importance of feedback and tangible interaction, however, there remain opportunities to incorporate real-time perception for adaptive assembly instructions\cite{cao_mobile_2023, nishihara_object_2015}


\textbf{Object Recognition in Augmented Reality and Assembly Tasks.} Integrating computer vision into AR has enabled more adaptive assembly workflows. Previous research has utilized object recognition to recognize assembly states, automatically advance AR instructions \cite{stanescu_state-aware_2023}, align virtual model registration \cite{yan_augmented_2022}, and verify whether a part has reached its target placement location \cite{kastner_integrative_2020}. Outside of AR-assisted assembly, vision–language models have also been used to detect assembly states in robotic assembly. However, pretrained models typically perform well only on common assembly tools and components they were trained on \cite{li_vlm-msgraph_2025}. In this research, we integrate object recognition with an automatic synthetic-data generation and training pipeline to enable learning on custom assembly components such as 3D-printed parts. Object recognition is then used to detect the real-time locations of individual parts and associate them with corresponding digital instructions by dynamically highlighting both their current and target spatial regions.

\textbf{Hand Tracking for Augmented Reality and Error Detection.}  
A growing body of work has investigated gesture-based AR interfaces \cite{gavgiotaki_gesture-based_2023, kyaw_gesture_2024}. Gestural recognition has been used for feedback-based assembly instructions, but primarily as an input to update the step sequence or visual display \cite{kyaw_humanmachine_2024,spencer_extended_2023}. Recent systems has combine hand tracking/action recognition with object detection to verify stage completion and detect errors \cite{zhang_learning-based_2025, mokuwe_assembly_2022}. However, this is primarily implemented as a quality-control step using fixed-camera setups and computer vision, rather than paired with augmented–reality–based visual instruction displays. In this research, we integrate hand and object tracking with AR instructions to enable real-time visual guidance and error detection. Moreover, instead of merely detecting errors, our system dynamically updates the assembly instructions based on the identified errors. 

\textbf{Synthesis.} While prior systems have used object recognition to identify assembly states, advance instructions, or align virtual models, our approach extends this by visually highlighting the real-time location of individual components in the physical workspace and indicating their corresponding target positions in the AR display. Although previous work has used hand tracking in AR primarily for input, our approach combines hand and object tracking within an AR environment to provide real-time error detection and adaptive instruction updates. Instead of treating errors as deviations to be corrected, our system interprets them as opportunities for creative divergence, allowing assembly instructions to dynamically adjust to user intent or improvisation. This adaptive model further supports simulation and iteration for assessing structural stability in evolving assemblies.


\section{Methods} 

\begin{figure*} [h]
  \centering
  \includegraphics[width=1\linewidth]
   {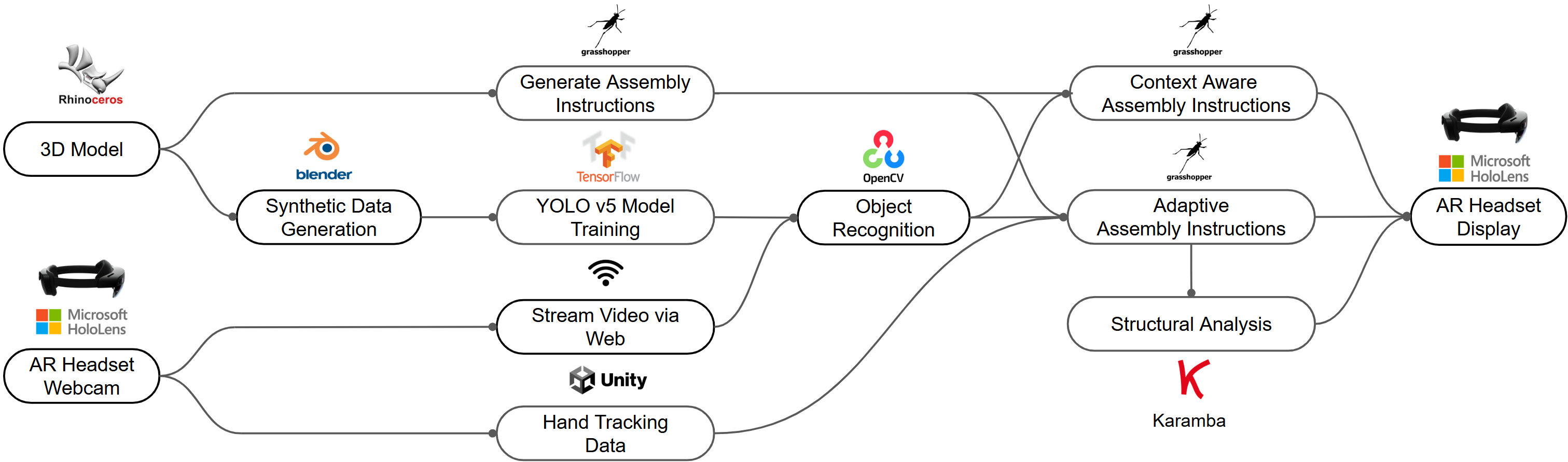}
  \caption{System pipeline diagram illustrating the inputs, software components, and data flow}
  \Description{System overview diagram illustrating the various software components and dataflow of the system}
  \label{workflow}
\end{figure*}

The system takes input from the HoloLens 2, including camera data for object recognition and hand-tracking data for error detection. The system also requires an assembly CAD model, which is used to derive the assembly instructions and sequence, as well as to extract individual component geometries for generating synthetic data to train the object recognition model. 
In this study, we present two demonstrations. The first uses LEGO components with a static 3D model to illustrate the baseline functionality of the AR instruction system with object recognition. The second uses 3D-printed components based on a parametric model modeled with predefined goals, integrating both object and hand tracking to demonstrate real-time error detection and adaptive instruction updates. In this setup, the system detects when users intentionally deviate from the prescribed sequence and automatically updates the assembly instructions to reflect the new configuration based on the parametric model. Additionally, the framework allows the user to see mid-assembly structural simulation to evaluate the stability of the updated configuration. See Figure \ref{workflow} and paragraphs below for more details about the dataflow within in the system pipeline. All computations are performed on a GPU-enabled workstation equipped with an NVIDIA RTX 3060 (CUDA support) and a multi-core Intel i9 CPU for model training and real-time execution.

\subsection{Object Recognition and Digital Twin of Physical Components in AR}
\begin{figure*} [h]
  \centering
  \includegraphics[width=1\linewidth]
   {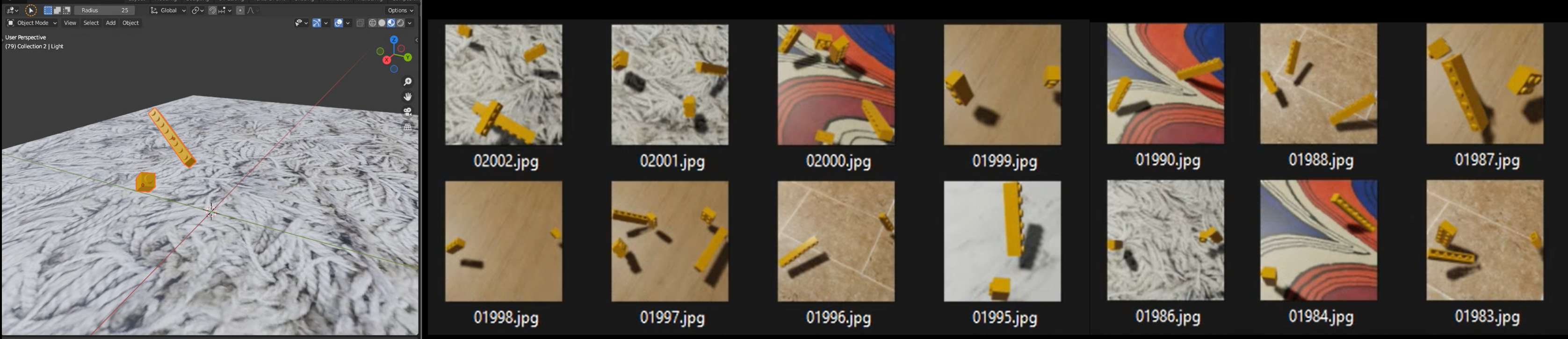}
  \caption{Automated synthetic data generation pipeline in Blender for training a deep learning-based object recognition model}
  \Description{Synthetic data generation setup for training a deep learning based object recognition model}
  \label{synthdata}
\end{figure*}

The object recognition algorithm is trained on synthetic data using the YOLOv5 (You Only Look Once) model. We developed a structured pipeline for generating a synthetic dataset under varied lighting, occlusion, and viewpoint conditions. A modular Blender scene was designed to simulate assembly environments \ref{synthdata}. We generate directional, point, and area lights with randomized intensity (200–1000 lux), color temperature (3000–6500 K), position, and direction for random light conditions. Multiple units are randomly placed and rotated in the scene to simulate partial overlaps and clutter. Azimuth (−180° to 180°) and elevation (15°–75°) were randomized to emulate diverse user perspectives. After setting the scene, we render the image using a path-tracing renderer to create photo-realistic images with bounding box coordinates and class labels alongside rendered images.

\begin{figure*} [h]
  \centering
  \includegraphics[width=1\linewidth]
   {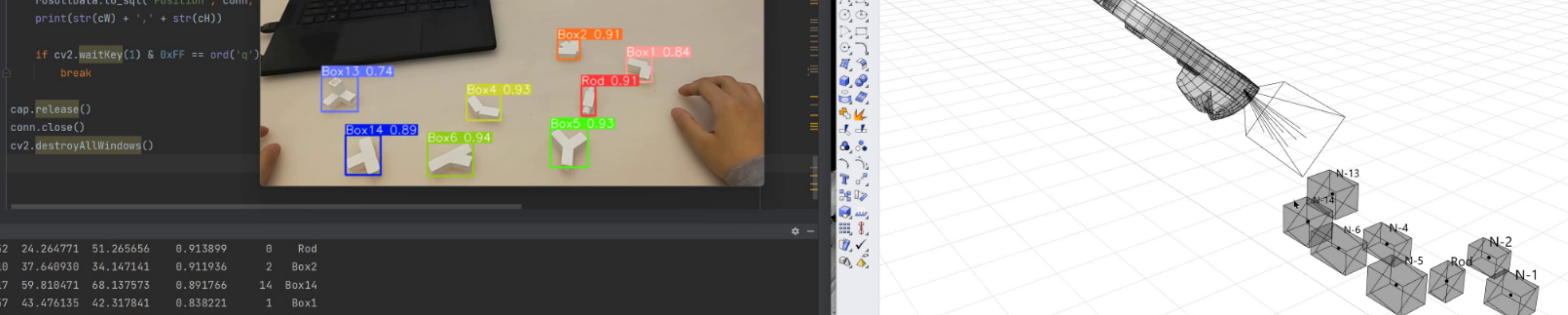}
  \caption{Digital twin of the physical assembly space, generated using 2D video-based object recognition and projected into a 3D planar representation.}
  \Description{Digital Twin}
  \label{digitaltwin}
\end{figure*}

We generated a dataset of approximately 9,000 synthetic images representing 8 types of bricks, and 16,000 images covering 15 types of custom 3D-printed components. The training set consists of synthetic images, while the validation and testing sets are composed of real images. We train a custom YOLOv5 object detection model, which is a single-stage architecture optimized for real-time inference, making it well-suited for latency-sensitive applications. The HoloLens 2 camera captures the physical workspace and streams video data to a server, where it is saved as individual frames and processed through the object recognition model. The model analyzes these frames to detect assembly components, outputting 2D bounding boxes and class IDs. These 2D bounding boxes are projected into the 3D AR environment using a homography-based 2D-to-3D planar projection computed from the camera’s pose and field of view, creating a digital twin of the physical components (Figure \ref{digitaltwin}. This involves aligning the HoloLens camera’s frustum with the 3D coordinate system, enabling real-time projection of detected assembly components in the AR interface. 

\subsection{Step by Step Assembly Instruction Generation and Visualization in AR}
\begin{figure*} [h]
  \centering
  \includegraphics[width=0.98\linewidth]
   {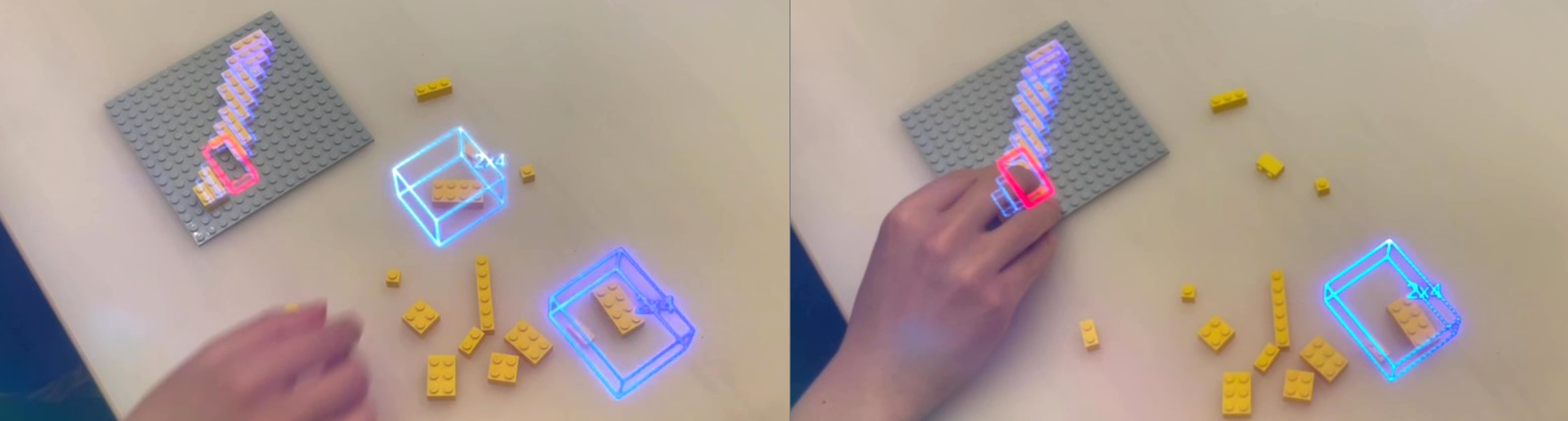}
  \caption{The interface guides users with step-by-step instructions, showing where to pick up and place each component.}
  \Description{The interface guides users with step-by-step instructions, showing where to pick up and place each component.}
  \label{pick}
\end{figure*}

\begin{figure*} [h]
  \centering
  \includegraphics[width=0.98\linewidth]
   {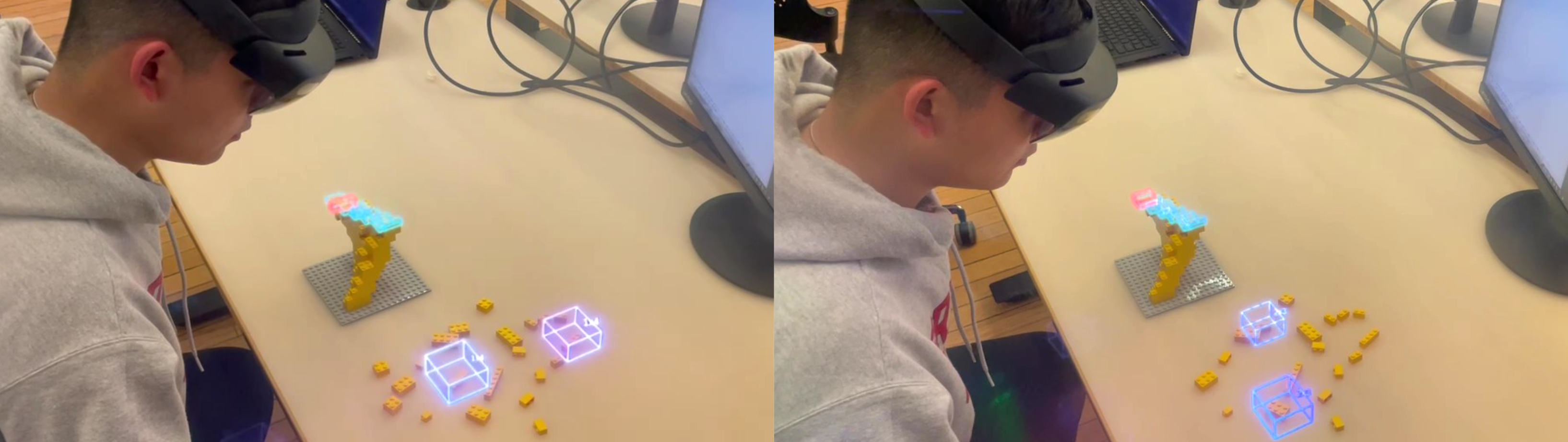}
  \caption{Assembly instructions for a twist wall LEGO assembly, connecting physical components locations with digital instructions.}
  \Description{Assembly instructions for a twist wall LEGO assembly, connecting physical components locations with digital instructions.}
  \label{steps}
\end{figure*}

The interface displays step-by-step assembly instructions, indicating where the user should pick up each component and where to place (Figure. \ref{pick}). For each instruction step, a corresponding 3D bounding box is generated to visually link the digital instruction with the physical component, highlighting both its current position and its intended placement (Figure. \ref{steps}). 3D bounding boxes are shown only for components relevant to the current assembly step. The assembly instruction sequence for the LEGO assembly is organized layer by layer. The assembly instructions for the 3D-printed components are further sorted by graph-based adjacency to ensure connectivity. The generation and sorting of assembly instructions are implemented using Rhino 3D’s CAD engine with Grasshopper for parametric scripting.

\subsection{Hand Tracking for Detecting User Interactions, Errors and Creative Deviations}

We use hand-tracking data from the HoloLens and object-recognition data from the YOLOv5 model to validate user actions. When the user selects the correct component, their hand enters the bounding box corresponding to the assembly step displayed in the interface. Upon correct selection, a green checkmark appears at the center of the component. Conversely, selecting the wrong component causes the user’s hand to intersect with a bounding box of components that are not relevant to the current assembly step and are not displayed. Upon incorrect selection, a red cross appears at the component. The system also tracks where the user places each component by monitoring the intersection of their hand with the bounding boxes of potential placement regions in the assembly.  In this paper, the error-detection feedback is implemented only for the 3D-printed components in our demonstration and not for the LEGO assembly task.



\subsection{Adaptive Instructions with Parametric Assembly Model and Structural Analysis}

\begin{figure} [h]
 \centering
 \includegraphics[width=\linewidth]
  {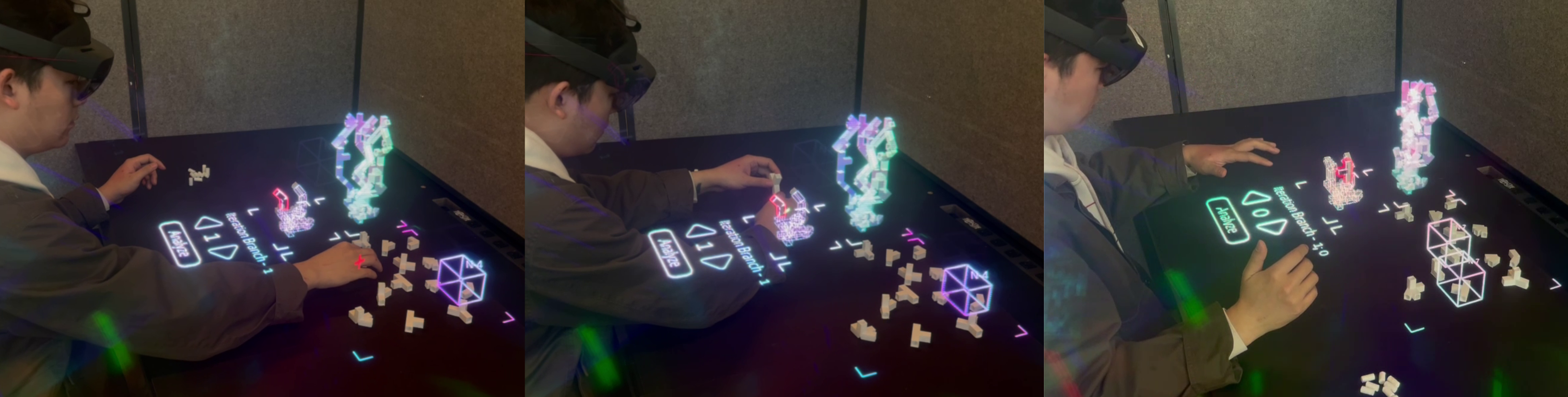}
 \caption{When the user picks and places the wrong component, the system detects the user deviation and adapts the assembly instruction to reflect the new assembly state and deviation.}
 \Description{When the user picks and places the wrong component, the system detects the user deviation and adapts the assembly instruction to reflect the new assembly state and deviation.}
 \label{error}
\end{figure}

\begin{figure} [h]
 \centering
 \includegraphics[width=\linewidth]
  {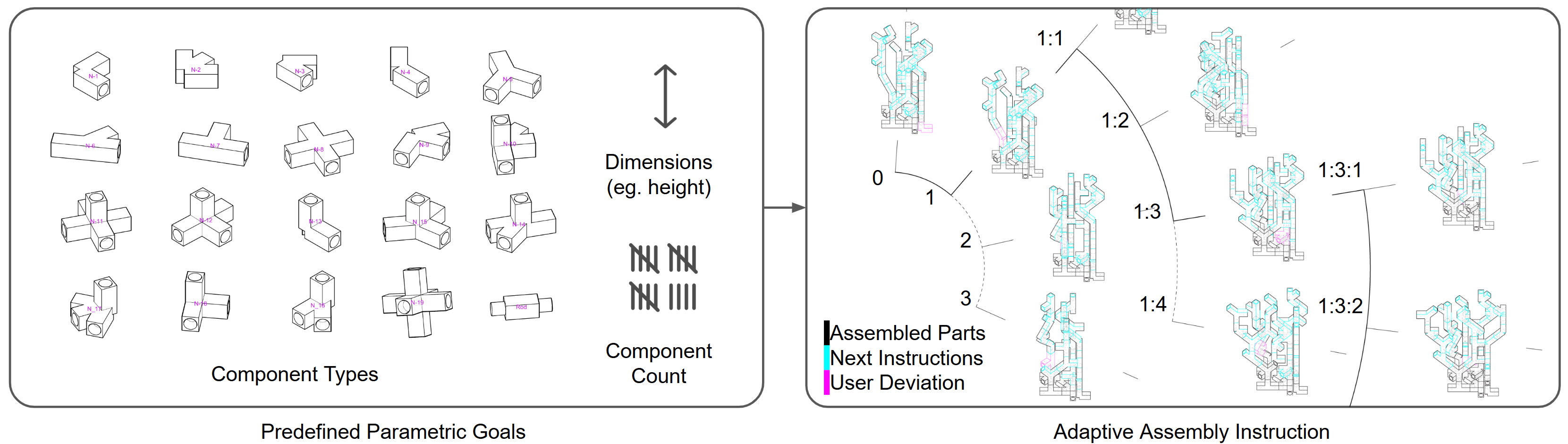}
 \caption{Parametric model with constraints and goals that helps determine possible structural iterations after assembly deviations.}
 \Description{Parametric model with constraints and goals that helps determine possible structural iterations after assembly deviations.}
 \label{iteration}
\end{figure}

When a user selects a different component or places a part in a location different from the original assembly instruction, the system adapts the assembly instruction based on the new assembly state (Figure. \ref{error}). The adaptive logic is driven by a parametric assembly model with predefined aggregation rules that specify how the 3D-printed nodal components can connect to one another, along with assembly goals such as satisfying the predetermined height of the structure and the limit on the number of components it can contain. When a deviation occurs, the parametric model triggers a constraint-based optimization process to re-evaluate possible configurations (Figure. \ref{iteration}). Using a graph-based exploration, it searches the design space to identify alternative assembly paths that both minimize deviation from the user’s current structure and maintain the intended design objectives. 

\begin{figure} [h]
 \centering
 \includegraphics[width=\linewidth]
  {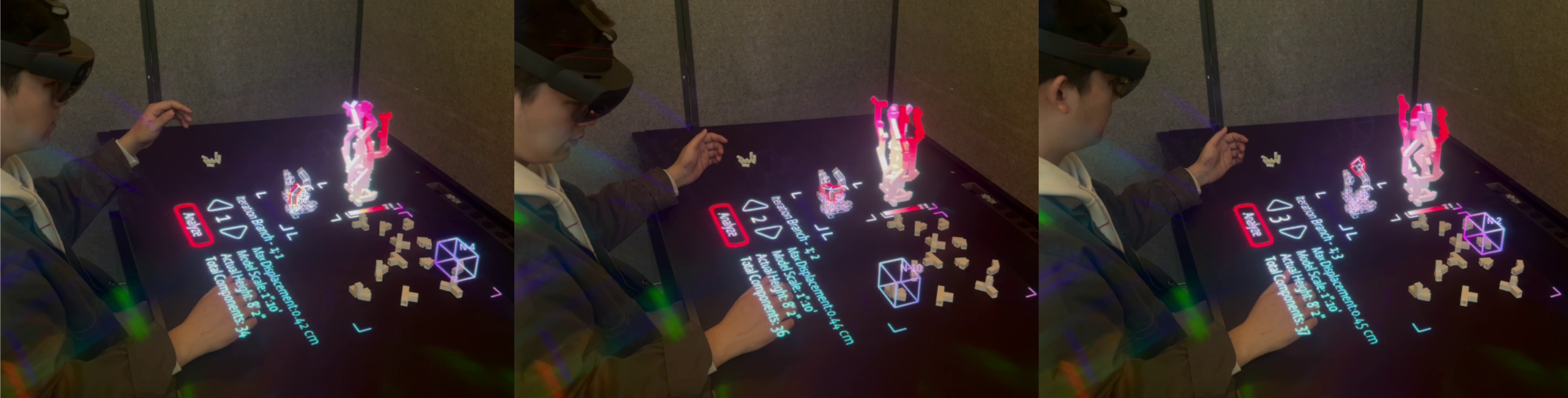}
 \caption{The user views the structural analysis and explores multiple assembly iterations based on the already assembled parts.}
 \label{structure}
\end{figure}

The interface allows the user to select through multiple potential assembly iterations that build upon the current state. These variations allow users to preview how different design trajectories might unfold based on what has already been assembled. This mechanism transforms error recovery into an opportunity for creative exploration, enabling users to choose among several viable directions rather than being constrained to a single predefined sequence \ref{structure}.  Additionally, the system integrates structural analysis by simulating factors based on gravity and the weight distribution of components baked into the parametric model. Users can view the corresponding structural performance for each assembly iteration in real time, supporting both stability and aesthetic exploration throughout the assembly process. Rather than treating the deviation as an error, the system interprets it as a creative exploration and updates the instruction accordingly

\section{Initial Results and Discussions}
\begin{figure} [h]
 \centering
 \includegraphics[width=\linewidth]
  {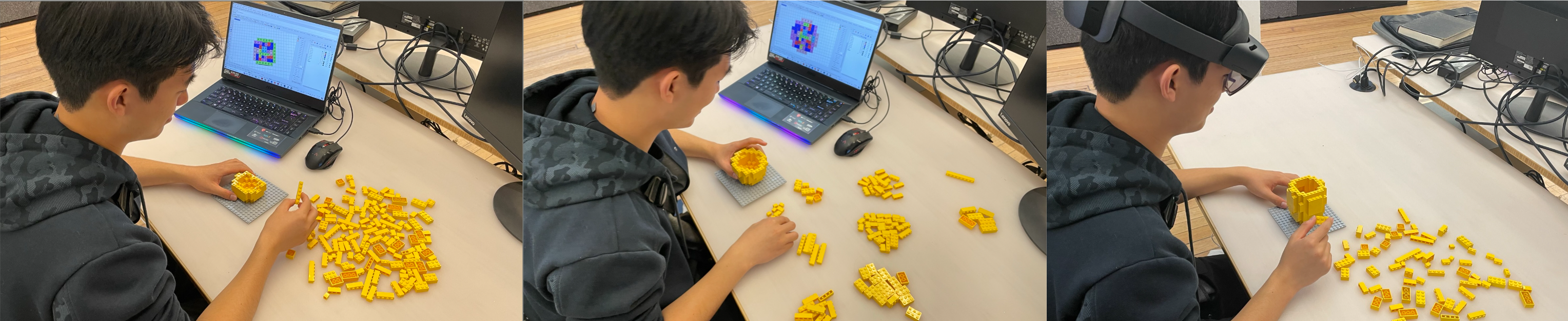}
 \caption{Assembly times for different instruction methods (left to right): unsorted (32:11), sorted (25:54), and AR-guided with object recognition (12:39) for the ellipsoid shaped LEGO assembly }
 \Description{Comparing Assembly time}
 \label{time}
\end{figure}

We conducted two demonstrations to evaluate the system’s capability to provide adaptive and responsive assembly guidance. The first demonstration used LEGO components to validate the baseline functionality of the AR assembly instruction system, focusing on object recognition accuracy, instruction visualization, and latency between physical interaction and digital feedback.  We demonstrated the assembly of two complex LEGO sculptures: an ellipsoidal egg and a twisted wall. Using our AR-guided system, we successfully assembled both structures without referencing any additional 2D paper drawings or 3D digital models, eliminating the need for manual searching, sorting, or labeling of parts. The demonstrations were designed as formative self-studies to compare the AR-guided system with object recognition, which took 12 minutes and 39 seconds, against the sorted assembly (25 minutes and 54 seconds) and unsorted assembly (32 minutes and 11 seconds). See figure \ref{time}. As this is a work-in-progress paper, future work includes a larger user study assessing effectiveness and usability qualitatively and quantitatively.

\begin{figure} [h]
 \centering
 \includegraphics[width=\linewidth]
  {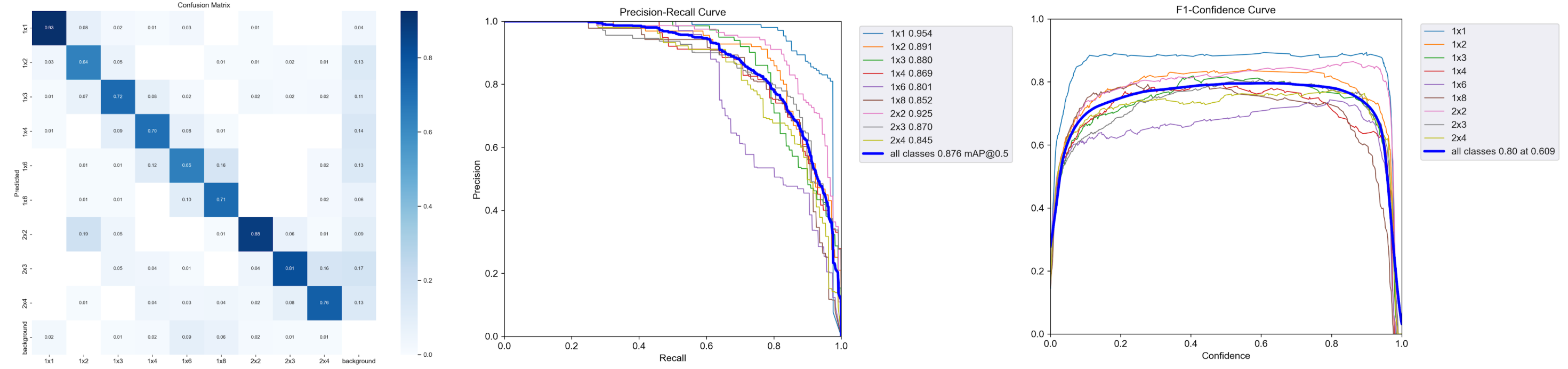}
 \caption{Confusion matrix, precision-recall (0.876), and F1-confidence (0.80) for the object recognition model identifying LEGOs blocks.}
 \Description{Sequence of photos of the chain executing planar folds; some joints approach torque limits.}
 \label{results}
\end{figure}

The custom YOLOv5 model trained on synthetic data for recognizing LEGO components achieved a mean average precision of 0.876 across all classes in the precision-recall curve, which is more than sufficient for real-time guidance. The F1–Confidence curves further confirm the model’s reliability, with an average F1 score of 0.80, indicating that the model not only detects the correct pieces but also misses very few (Figure. \ref{results}). Although trained primarily on synthetic images, the model can generalized to real-world scenes under varied lighting and occlusion, suggesting that the synthetic data pipeline enables accurate, real-time component recognition for AR guidance and digital twin construction.  While the current implementation is limited to tabletop assemblies using 2D–3D planar projection for object recognition, the system can scale to larger and more complex assembly tasks in future studies.

The second demonstration employed custom 3D-printed nodal components designed with a parametric assembly model.  Hand tracking allowed precise detection of user interactions with physical components. Across 25 assembly sequences, correct selections triggered positive confirmation with a detection accuracy of 96.2\%, while incorrect selections were accurately flagged with a red cross in 91.8\% of cases. In the 3D-printed assembly demonstration, deviations were introduced in the study by selecting components out of order or placing them in alternative orientations. The parametric model’s constraint-solving process consistently produced structurally valid alternatives that maintained the intended height and component limits. The updated assembly instructions successfully reconfigured the assembly path in all attempts, updating the AR guidance to reflect the new state. While this study only considers component types, height dimensions, and the number of components as its set parameters and goals, future studies could incorporate additional constraints or objectives or incorporate 3D generative AI in the making process \cite{kyaw_speech_2025, kyaw_making_2025}

Users can explore multiple alternative iterations through the parametric assembly model, demonstrating the potential of adaptive updates to foster curiosity and creative engagement. The demonstrations suggest that adaptive assembly guidance not only helps recover from errors but also transforms them into productive opportunities for exploration and iteration. Furthermore, integrating structural simulation into the assembly workflow provides users with immediate insight into stability during the process, highlighting its potential to support iterative design decisions and informed creative exploration by bridging hands-on making and computational feedback \cite{kyaw_active_2023}. This approach could be also extended to larger-scale AR assembly \cite{kyaw_augmented_2023}, where adaptive guidance and real-time feedback would be critical for managing complexity, ensuring structural integrity, and enabling collaborative construction workflows.

The results demonstrate the potential of integrating perception-driven feedback loops into AR-guided assembly. While traditional AR systems follow fixed sequences, our framework dynamically interprets user-embodied interaction with physical components in the assembly space and adapts instructions accordingly. The coupling of object recognition, hand tracking, and parametric modeling creates a bidirectional relationship between human action and digital feedback, transforming AR guidance from a passive instruction display into an active collaborator in the physical making process. 


\section{GenAI Usage Disclosure}
We used ChatGPT 5 (OpenAI, 2025) for minor grammar and language editing of author-written text. We did not use AI tools for idea generation, figures, data analysis or code generation.



\bibliographystyle{ACM-Reference-Format}
\bibliography{references}


\begin{thebibliography}{33}


\ifx \showCODEN    \undefined \def \showCODEN     #1{\unskip}     \fi
\ifx \showISBNx    \undefined \def \showISBNx     #1{\unskip}     \fi
\ifx \showISBNxiii \undefined \def \showISBNxiii  #1{\unskip}     \fi
\ifx \showISSN     \undefined \def \showISSN      #1{\unskip}     \fi
\ifx \showLCCN     \undefined \def \showLCCN      #1{\unskip}     \fi
\ifx \shownote     \undefined \def \shownote      #1{#1}          \fi
\ifx \showarticletitle \undefined \def \showarticletitle #1{#1}   \fi
\ifx \showURL      \undefined \def \showURL       {\relax}        \fi
\providecommand\bibfield[2]{#2}
\providecommand\bibinfo[2]{#2}
\providecommand\natexlab[1]{#1}
\providecommand\showeprint[2][]{arXiv:#2}

\bibitem[Alves et~al\mbox{.}(2022)]%
        {alves_comparing_2022}
\bibfield{author}{\bibinfo{person}{João Alves}, \bibinfo{person}{Bernardo Marques}, \bibinfo{person}{Carlos Ferreira}, \bibinfo{person}{Paulo Dias}, {and} \bibinfo{person}{Beatriz Santos}.} \bibinfo{year}{2022}\natexlab{}.
\newblock \showarticletitle{Comparing augmented reality visualization methods for assembly procedures}.
\newblock \bibinfo{journal}{\emph{Virtual Reality}}  \bibinfo{volume}{26} (\bibinfo{date}{March} \bibinfo{year}{2022}), \bibinfo{pages}{1--14}.
\newblock
\href{https://doi.org/10.1007/s10055-021-00557-8}{doi:\nolinkurl{10.1007/s10055-021-00557-8}}


\bibitem[Cao et~al\mbox{.}(2023)]%
        {cao_mobile_2023}
\bibfield{author}{\bibinfo{person}{Jacky Cao}, \bibinfo{person}{Kit-Yung Lam}, \bibinfo{person}{Lik-Hang Lee}, \bibinfo{person}{Xiaoli Liu}, \bibinfo{person}{Pan Hui}, {and} \bibinfo{person}{Xiang Su}.} \bibinfo{year}{2023}\natexlab{}.
\newblock \showarticletitle{Mobile {Augmented} {Reality}: {User} {Interfaces}, {Frameworks}, and {Intelligence}}.
\newblock \bibinfo{journal}{\emph{ACM Comput. Surv.}} \bibinfo{volume}{55}, \bibinfo{number}{9} (\bibinfo{date}{Jan.} \bibinfo{year}{2023}), \bibinfo{pages}{189:1--189:36}.
\newblock
\showISSN{0360-0300}
\href{https://doi.org/10.1145/3557999}{doi:\nolinkurl{10.1145/3557999}}


\bibitem[Chen et~al\mbox{.}(2024)]%
        {chen_augmenting_2024}
\bibfield{author}{\bibinfo{person}{Zhenfang Chen}, \bibinfo{person}{Tate Johnson}, \bibinfo{person}{Andrew Knowles}, \bibinfo{person}{Ann Li}, \bibinfo{person}{Semina Yi}, \bibinfo{person}{Yumeng Zhuang}, \bibinfo{person}{Daragh Byrne}, {and} \bibinfo{person}{Dina El-Zanfaly}.} \bibinfo{year}{2024}\natexlab{}.
\newblock \showarticletitle{Augmenting {Embodied} {Learning} in {Welding} {Training}: {The} {Co}-{Design} of an {XR}- and {tinyML}-{Enabled} {Welding} {System} for {Creative} {Arts} and {Manufacturing} {Training}}. In \bibinfo{booktitle}{\emph{Proceedings of the {Eighteenth} {International} {Conference} on {Tangible}, {Embedded}, and {Embodied} {Interaction}}} \emph{(\bibinfo{series}{{TEI} '24})}. \bibinfo{publisher}{Association for Computing Machinery}, \bibinfo{address}{New York, NY, USA}, \bibinfo{pages}{1--14}.
\newblock
\showISBNx{9798400704024}
\href{https://doi.org/10.1145/3623509.3633398}{doi:\nolinkurl{10.1145/3623509.3633398}}


\bibitem[Fang et~al\mbox{.}(2023)]%
        {fang_research_2023-1}
\bibfield{author}{\bibinfo{person}{Wei Fang}, \bibinfo{person}{Shuhong Xu}, \bibinfo{person}{Lei Han}, {and} \bibinfo{person}{Zhangwenchi Li}.} \bibinfo{year}{2023}\natexlab{}.
\newblock \showarticletitle{Research and {Application} {Progress} of {Tracking} {Registration} {Methods} in {AR} {Assembly}}.
\newblock \bibinfo{journal}{\emph{Journal of System Simulation}} \bibinfo{volume}{35}, \bibinfo{number}{7} (\bibinfo{date}{Aug.} \bibinfo{year}{2023}), \bibinfo{pages}{1438--1454}.
\newblock
\showISSN{1004-731X}
\href{https://doi.org/10.16182/j.issn1004731x.joss.22-0335}{doi:\nolinkurl{10.16182/j.issn1004731x.joss.22-0335}}


\bibitem[Funk et~al\mbox{.}(2017)]%
        {funk_working_2017}
\bibfield{author}{\bibinfo{person}{Markus Funk}, \bibinfo{person}{Andreas Bächler}, \bibinfo{person}{Liane Bächler}, \bibinfo{person}{Thomas Kosch}, \bibinfo{person}{Thomas Heidenreich}, {and} \bibinfo{person}{Albrecht Schmidt}.} \bibinfo{year}{2017}\natexlab{}.
\newblock \bibinfo{booktitle}{\emph{Working with {Augmented} {Reality}?: {A} {Long}-{Term} {Analysis} of {In}-{Situ} {Instructions} at the {Assembly} {Workplace}}}.
\newblock
\href{https://doi.org/10.1145/3056540.3056548}{doi:\nolinkurl{10.1145/3056540.3056548}}
\newblock
\shownote{Pages: 229}.


\bibitem[Garcia~de Paiva~Pinheiro and Cuperschmid(2025)]%
        {garcia_de_paiva_pinheiro_integrating_2025}
\bibfield{author}{\bibinfo{person}{Ana~Ester Garcia~de Paiva~Pinheiro} {and} \bibinfo{person}{Ana~Regina Cuperschmid}.} \bibinfo{year}{2025}\natexlab{}.
\newblock \showarticletitle{Integrating {Augmented} {Reality} and {Artificial} {Intelligence} in {Assembly} {Tasks}: {A} {Review} of {Strategies}, {Tools}, and {Challenges}}.
\newblock  (\bibinfo{date}{Aug.} \bibinfo{year}{2025}).
\newblock
\href{https://doi.org/10.6084/m9.figshare.29881853.v1}{doi:\nolinkurl{10.6084/m9.figshare.29881853.v1}}
\newblock
\shownote{Publisher: figshare}.


\bibitem[Gavgiotaki et~al\mbox{.}(2023)]%
        {gavgiotaki_gesture-based_2023}
\bibfield{author}{\bibinfo{person}{Despoina Gavgiotaki}, \bibinfo{person}{Stavroula Ntoa}, \bibinfo{person}{George Margetis}, \bibinfo{person}{Konstantinos~C. Apostolakis}, {and} \bibinfo{person}{Constantine Stephanidis}.} \bibinfo{year}{2023}\natexlab{}.
\newblock \showarticletitle{Gesture-based {Interaction} for {AR} {Systems}: {A} {Short} {Review}}. In \bibinfo{booktitle}{\emph{Proceedings of the 16th {International} {Conference} on {PErvasive} {Technologies} {Related} to {Assistive} {Environments}}} \emph{(\bibinfo{series}{{PETRA} '23})}. \bibinfo{publisher}{Association for Computing Machinery}, \bibinfo{address}{New York, NY, USA}, \bibinfo{pages}{284--292}.
\newblock
\showISBNx{9798400700699}
\href{https://doi.org/10.1145/3594806.3594815}{doi:\nolinkurl{10.1145/3594806.3594815}}


\bibitem[Han et~al\mbox{.}(2025)]%
        {han_immersive_2025}
\bibfield{author}{\bibinfo{person}{Teresa Han}, \bibinfo{person}{Qi Wang}, \bibinfo{person}{Zhiyong Dong}, \bibinfo{person}{Peter Búš}, \bibinfo{person}{Zhiqian Liu}, {and} \bibinfo{person}{Ruwei Chen}.} \bibinfo{year}{2025}\natexlab{}.
\newblock \showarticletitle{Immersive {AR}-assisted assemblies for self-building strategies using prefabricated timber components}.
\newblock \bibinfo{journal}{\emph{Architectural Intelligence}} \bibinfo{volume}{4}, \bibinfo{number}{1} (\bibinfo{date}{July} \bibinfo{year}{2025}), \bibinfo{pages}{12}.
\newblock
\showISSN{2731-6726}
\href{https://doi.org/10.1007/s44223-025-00091-6}{doi:\nolinkurl{10.1007/s44223-025-00091-6}}


\bibitem[Hattab and Taubin(2019)]%
        {hattab_interactive_2019}
\bibfield{author}{\bibinfo{person}{Ammar Hattab} {and} \bibinfo{person}{Gabriel Taubin}.} \bibinfo{year}{2019}\natexlab{}.
\newblock \showarticletitle{Interactive {Fabrication} of {CSG} {Models} with {Assisted} {Carving}}. In \bibinfo{booktitle}{\emph{Proceedings of the {Thirteenth} {International} {Conference} on {Tangible}, {Embedded}, and {Embodied} {Interaction}}} \emph{(\bibinfo{series}{{TEI} '19})}. \bibinfo{publisher}{Association for Computing Machinery}, \bibinfo{address}{New York, NY, USA}, \bibinfo{pages}{677--682}.
\newblock
\showISBNx{978-1-4503-6196-5}
\href{https://doi.org/10.1145/3294109.3295644}{doi:\nolinkurl{10.1145/3294109.3295644}}


\bibitem[Henderson and Feiner(2011)]%
        {henderson_augmented_2011}
\bibfield{author}{\bibinfo{person}{Steven~J. Henderson} {and} \bibinfo{person}{Steven~K. Feiner}.} \bibinfo{year}{2011}\natexlab{}.
\newblock \showarticletitle{Augmented reality in the psychomotor phase of a procedural task}. In \bibinfo{booktitle}{\emph{2011 10th {IEEE} {International} {Symposium} on {Mixed} and {Augmented} {Reality}}}. \bibinfo{pages}{191--200}.
\newblock
\href{https://doi.org/10.1109/ISMAR.2011.6092386}{doi:\nolinkurl{10.1109/ISMAR.2011.6092386}}


\bibitem[Kang et~al\mbox{.}(2018)]%
        {kang_prototyping_2018}
\bibfield{author}{\bibinfo{person}{Seokbin Kang}, \bibinfo{person}{Leyla Norooz}, \bibinfo{person}{Virginia Byrne}, \bibinfo{person}{Tamara Clegg}, {and} \bibinfo{person}{Jon~E. Froehlich}.} \bibinfo{year}{2018}\natexlab{}.
\newblock \showarticletitle{Prototyping and {Simulating} {Complex} {Systems} with {Paper} {Craft} and {Augmented} {Reality}: {An} {Initial} {Investigation}}. In \bibinfo{booktitle}{\emph{Proceedings of the {Twelfth} {International} {Conference} on {Tangible}, {Embedded}, and {Embodied} {Interaction}}} \emph{(\bibinfo{series}{{TEI} '18})}. \bibinfo{publisher}{Association for Computing Machinery}, \bibinfo{address}{New York, NY, USA}, \bibinfo{pages}{320--328}.
\newblock
\showISBNx{978-1-4503-5568-1}
\href{https://doi.org/10.1145/3173225.3173264}{doi:\nolinkurl{10.1145/3173225.3173264}}


\bibitem[Kyaw(2023)]%
        {kyaw_active_2023}
\bibfield{author}{\bibinfo{person}{Alexander~Htet Kyaw}.} \bibinfo{year}{2023}\natexlab{}.
\newblock \showarticletitle{Active {Bending} in {Physics}-{Based} {Mixed} {Reality}: {The} design and fabrication of a reconfigurable modular bamboo system}. In \bibinfo{booktitle}{\emph{Dokonal, {W}, {Hirschberg}, {U} and {Wurzer}, {G} (eds.), {Digital} {Design} {Reconsidered} - {Proceedings} of the 41st {Conference} on {Education} and {Research} in {Computer} {Aided} {Architectural} {Design} in {Europe} ({eCAADe} 2023) - {Volume} 1, {Graz}, 20-22 {September} 2023, pp. 169–178}}. \bibinfo{publisher}{CUMINCAD}.
\newblock
\urldef\tempurl%
\url{https://papers.cumincad.org/cgi-bin/works/paper/ecaade2023_447}
\showURL{%
\tempurl}


\bibitem[Kyaw et~al\mbox{.}(2023)]%
        {kyaw_augmented_2023}
\bibfield{author}{\bibinfo{person}{Alexander~Htet Kyaw}, \bibinfo{person}{Arvin HaoCheng~Xu}, \bibinfo{person}{Gwyllim Jahn}, \bibinfo{person}{Nick van~den Berg}, \bibinfo{person}{Cameron Newnham}, {and} \bibinfo{person}{Sasa Zivkovic}.} \bibinfo{year}{2023}\natexlab{}.
\newblock \showarticletitle{Augmented {Reality} for high precision fabrication of {Glued} {Laminated} {Timber} beams}.
\newblock \bibinfo{journal}{\emph{Automation in Construction}}  \bibinfo{volume}{152} (\bibinfo{date}{Aug.} \bibinfo{year}{2023}), \bibinfo{pages}{104912}.
\newblock
\showISSN{0926-5805}
\href{https://doi.org/10.1016/j.autcon.2023.104912}{doi:\nolinkurl{10.1016/j.autcon.2023.104912}}


\bibitem[Kyaw et~al\mbox{.}(2025a)]%
        {kyaw_making_2025}
\bibfield{author}{\bibinfo{person}{Alexander~Htet Kyaw}, \bibinfo{person}{Se~Hwan Jeon}, \bibinfo{person}{Miana Smith}, {and} \bibinfo{person}{Neil Gershenfeld}.} \bibinfo{year}{2025}\natexlab{a}.
\newblock \bibinfo{title}{Making {Physical} {Objects} with {Generative} {AI} and {Robotic} {Assembly}: {Considering} {Fabrication} {Constraints}, {Sustainability}, {Time}, {Functionality}, and {Accessibility}}.
\newblock
\href{https://doi.org/10.48550/arXiv.2504.19131}{doi:\nolinkurl{10.48550/arXiv.2504.19131}}
\newblock
\shownote{arXiv:2504.19131 [cs]}.


\bibitem[Kyaw et~al\mbox{.}(2025b)]%
        {kyaw_ai_2025}
\bibfield{author}{\bibinfo{person}{Alexander~Htet Kyaw}, \bibinfo{person}{Haotian Ma}, \bibinfo{person}{Sasa Zivkovic}, {and} \bibinfo{person}{Jenny Sabin}.} \bibinfo{year}{2025}\natexlab{b}.
\newblock \bibinfo{title}{{AI} {Assisted} {AR} {Assembly}: {Object} {Recognition} and {Computer} {Vision} for {Augmented} {Reality} {Assisted} {Assembly}}.
\newblock
\href{https://doi.org/10.48550/arXiv.2511.05394}{doi:\nolinkurl{10.48550/arXiv.2511.05394}}
\newblock
\shownote{arXiv:2511.05394 [cs]}.


\bibitem[Kyaw et~al\mbox{.}(2025c)]%
        {kyaw_speech_2025}
\bibfield{author}{\bibinfo{person}{Alexander~Htet Kyaw}, \bibinfo{person}{Miana Smith}, \bibinfo{person}{Se~Hwan Jeon}, {and} \bibinfo{person}{Neil Gershenfeld}.} \bibinfo{year}{2025}\natexlab{c}.
\newblock \bibinfo{title}{Speech to {Reality}: {On}-{Demand} {Production} using {Natural} {Language}, {3D} {Generative} {AI}, and {Discrete} {Robotic} {Assembly}}.
\newblock
\href{https://doi.org/10.1145/3745778.3766670}{doi:\nolinkurl{10.1145/3745778.3766670}}
\newblock
\shownote{arXiv:2409.18390 [cs]}.


\bibitem[Kyaw et~al\mbox{.}(2024a)]%
        {kyaw_humanmachine_2024}
\bibfield{author}{\bibinfo{person}{Alexander~Htet Kyaw}, \bibinfo{person}{Lawson Spencer}, {and} \bibinfo{person}{Leslie Lok}.} \bibinfo{year}{2024}\natexlab{a}.
\newblock \showarticletitle{Human–machine collaboration using gesture recognition in mixed reality and robotic fabrication}.
\newblock \bibinfo{journal}{\emph{Architectural Intelligence}} \bibinfo{volume}{3}, \bibinfo{number}{1} (\bibinfo{date}{March} \bibinfo{year}{2024}), \bibinfo{pages}{11}.
\newblock
\showISSN{2731-6726}
\href{https://doi.org/10.1007/s44223-024-00053-4}{doi:\nolinkurl{10.1007/s44223-024-00053-4}}


\bibitem[Kyaw et~al\mbox{.}(2024b)]%
        {kyaw_gesture_2024}
\bibfield{author}{\bibinfo{person}{Alexander~Htet Kyaw}, \bibinfo{person}{Lawson Spencer}, \bibinfo{person}{Sasa Zivkovic}, {and} \bibinfo{person}{Leslie Lok}.} \bibinfo{year}{2024}\natexlab{b}.
\newblock \bibinfo{booktitle}{\emph{Gesture {Recognition} for {Feedback} {Based} {Mixed} {Reality} and {Robotic} {Fabrication}: {A} {Case} {Study} of the {UnLog} {Tower}}}. Vol.~\bibinfo{volume}{Phygital Intelligence}.
\newblock
\showISBNx{978-981-9984-04-6}
\href{https://doi.org/10.1007/978-981-99-8405-3_28}{doi:\nolinkurl{10.1007/978-981-99-8405-3_28}}
\newblock
\shownote{Pages: 345}.


\bibitem[Kästner et~al\mbox{.}(2020)]%
        {kastner_integrative_2020}
\bibfield{author}{\bibinfo{person}{Linh Kästner}, \bibinfo{person}{Leon Eversberg}, \bibinfo{person}{Marina Mursa}, {and} \bibinfo{person}{Jens Lambrecht}.} \bibinfo{year}{2020}\natexlab{}.
\newblock \bibinfo{title}{Integrative {Object} and {Pose} to {Task} {Detection} for an {Augmented}-{Reality}-based {Human} {Assistance} {System} using {Neural} {Networks}}.
\newblock
\href{https://doi.org/10.48550/arXiv.2008.13419}{doi:\nolinkurl{10.48550/arXiv.2008.13419}}
\newblock
\shownote{arXiv:2008.13419 [cs]}.


\bibitem[Lee et~al\mbox{.}(2023)]%
        {lee_design_2023}
\bibfield{author}{\bibinfo{person}{Benjamin Lee}, \bibinfo{person}{Michael Sedlmair}, {and} \bibinfo{person}{Dieter Schmalstieg}.} \bibinfo{year}{2023}\natexlab{}.
\newblock \showarticletitle{Design {Patterns} for {Situated} {Visualization} in {Augmented} {Reality}}.
\newblock \bibinfo{journal}{\emph{IEEE Transactions on Visualization and Computer Graphics}} (\bibinfo{year}{2023}), \bibinfo{pages}{1--12}.
\newblock
\showISSN{1077-2626, 1941-0506, 2160-9306}
\href{https://doi.org/10.1109/TVCG.2023.3327398}{doi:\nolinkurl{10.1109/TVCG.2023.3327398}}
\newblock
\shownote{arXiv:2307.09157 [cs]}.


\bibitem[Lee-Cultura and Giannakos(2020)]%
        {lee-cultura_embodied_2020}
\bibfield{author}{\bibinfo{person}{Serena Lee-Cultura} {and} \bibinfo{person}{Michail Giannakos}.} \bibinfo{year}{2020}\natexlab{}.
\newblock \showarticletitle{Embodied {Interaction} and {Spatial} {Skills}: {A} {Systematic} {Review} of {Empirical} {Studies}}.
\newblock \bibinfo{journal}{\emph{Interacting with Computers}} \bibinfo{volume}{32}, \bibinfo{number}{4} (\bibinfo{date}{July} \bibinfo{year}{2020}), \bibinfo{pages}{331--366}.
\newblock
\showISSN{1873-7951}
\href{https://doi.org/10.1093/iwcomp/iwaa023}{doi:\nolinkurl{10.1093/iwcomp/iwaa023}}


\bibitem[Li et~al\mbox{.}(2025)]%
        {li_vlm-msgraph_2025}
\bibfield{author}{\bibinfo{person}{Shufei Li}, \bibinfo{person}{Zhijie Yan}, \bibinfo{person}{Zuoxu Wang}, {and} \bibinfo{person}{Yiping Gao}.} \bibinfo{year}{2025}\natexlab{}.
\newblock \showarticletitle{{VLM}-{MSGraph}: {Vision} {Language} {Model}-enabled {Multi}-hierarchical {Scene} {Graph} for robotic assembly}.
\newblock \bibinfo{journal}{\emph{Robotics and Computer-Integrated Manufacturing}}  \bibinfo{volume}{94} (\bibinfo{date}{Aug.} \bibinfo{year}{2025}), \bibinfo{pages}{102978}.
\newblock
\showISSN{0736-5845}
\href{https://doi.org/10.1016/j.rcim.2025.102978}{doi:\nolinkurl{10.1016/j.rcim.2025.102978}}


\bibitem[Liu et~al\mbox{.}(2023)]%
        {liu_instrumentar_2023}
\bibfield{author}{\bibinfo{person}{Ziyi Liu}, \bibinfo{person}{Zhengzhe Zhu}, \bibinfo{person}{Enze Jiang}, \bibinfo{person}{Feichi Huang}, \bibinfo{person}{Ana~M Villanueva}, \bibinfo{person}{Xun Qian}, \bibinfo{person}{Tianyi Wang}, {and} \bibinfo{person}{Karthik Ramani}.} \bibinfo{year}{2023}\natexlab{}.
\newblock \showarticletitle{{InstruMentAR}: {Auto}-{Generation} of {Augmented} {Reality} {Tutorials} for {Operating} {Digital} {Instruments} {Through} {Recording} {Embodied} {Demonstration}}. In \bibinfo{booktitle}{\emph{Proceedings of the 2023 {CHI} {Conference} on {Human} {Factors} in {Computing} {Systems}}} \emph{(\bibinfo{series}{{CHI} '23})}. \bibinfo{publisher}{Association for Computing Machinery}, \bibinfo{address}{New York, NY, USA}, \bibinfo{pages}{1--17}.
\newblock
\showISBNx{978-1-4503-9421-5}
\href{https://doi.org/10.1145/3544548.3581442}{doi:\nolinkurl{10.1145/3544548.3581442}}


\bibitem[Ma and Kyaw(2023)]%
        {ma_ai_2023}
\bibfield{author}{\bibinfo{person}{Haotian Ma} {and} \bibinfo{person}{Alexander~Htet Kyaw}.} \bibinfo{year}{2023}\natexlab{}.
\newblock \showarticletitle{{AI} {ASSEMBLY}: {OBJECT} {RECOGNITION}, {COMPUTER} {VISION}, {AND} {DIGITAL} {TWIN} {FOR} {MIXED} {REALITY} {ASSEMBLY}}.
\newblock  (\bibinfo{date}{May} \bibinfo{year}{2023}).
\newblock
\urldef\tempurl%
\url{https://hdl.handle.net/1813/113919}
\showURL{%
\tempurl}


\bibitem[Mokuwe et~al\mbox{.}(2022)]%
        {mokuwe_assembly_2022}
\bibfield{author}{\bibinfo{person}{Mamuku Mokuwe}, \bibinfo{person}{Yurisha Goorun}, {and} \bibinfo{person}{Gerrie Crafford}.} \bibinfo{year}{2022}\natexlab{}.
\newblock \showarticletitle{Assembly {Line} {Quality} {Assurance} {Through} {Hand} {Tracking} and {Object} {Detection}}.
\newblock \bibinfo{journal}{\emph{MATEC Web of Conferences}}  \bibinfo{volume}{370} (\bibinfo{year}{2022}), \bibinfo{pages}{07003}.
\newblock
\showISSN{2261-236X}
\href{https://doi.org/10.1051/matecconf/202237007003}{doi:\nolinkurl{10.1051/matecconf/202237007003}}


\bibitem[Nishihara and Okamoto(2015)]%
        {nishihara_object_2015}
\bibfield{author}{\bibinfo{person}{Anderson Nishihara} {and} \bibinfo{person}{Jun Okamoto}.} \bibinfo{year}{2015}\natexlab{}.
\newblock \showarticletitle{Object recognition in assembly assisted by augmented reality system}. In \bibinfo{booktitle}{\emph{2015 {SAI} {Intelligent} {Systems} {Conference} ({IntelliSys})}}. \bibinfo{pages}{400--407}.
\newblock
\href{https://doi.org/10.1109/IntelliSys.2015.7361172}{doi:\nolinkurl{10.1109/IntelliSys.2015.7361172}}


\bibitem[Spencer(2023)]%
        {spencer_extended_2023}
\bibfield{author}{\bibinfo{person}{Lawson; Htet~Kyaw Spencer}.} \bibinfo{year}{2023}\natexlab{}.
\newblock \showarticletitle{Extended {Reality} {Workflows} for {Multi}-{Material} {Construction} and {Assemblies}}. In \bibinfo{booktitle}{\emph{{ACADIA} 2023: {Habits} of the {Anthropocene}: {Scarcity} and {Abundance} in a {Post}-{Material} {Economy} [{Volume} 2: {Proceedings} of the 43rd {Annual} {Conference} for the {Association} for {Computer} {Aided} {Design} in {Architecture} ({ACADIA}) {ISBN} 979-8-9891764-0-3]. {Denver}. 26-28 {October} 2023. edited by {A}. {Crawford}, {N}. {Diniz}, {R}. {Beckett}, {J}. {Vanucchi}, {M}. {Swackhamer} 318-328.}} \bibinfo{publisher}{CUMINCAD}.
\newblock
\urldef\tempurl%
\url{https://papers.cumincad.org/cgi-bin/works/paper/acadia23_v2_318}
\showURL{%
\tempurl}


\bibitem[Stanescu et~al\mbox{.}(2023)]%
        {stanescu_state-aware_2023}
\bibfield{author}{\bibinfo{person}{Ana Stanescu}, \bibinfo{person}{Peter Mohr}, \bibinfo{person}{Mateusz Kozinski}, \bibinfo{person}{Shohei Mori}, \bibinfo{person}{Dieter Schmalstieg}, {and} \bibinfo{person}{Denis Kalkofen}.} \bibinfo{year}{2023}\natexlab{}.
\newblock \showarticletitle{State-{Aware} {Configuration} {Detection} for {Augmented} {Reality} {Step}-by-{Step} {Tutorials}}. In \bibinfo{booktitle}{\emph{2023 {IEEE} {International} {Symposium} on {Mixed} and {Augmented} {Reality} ({ISMAR})}}. \bibinfo{publisher}{IEEE}, \bibinfo{address}{Sydney, Australia}, \bibinfo{pages}{157--166}.
\newblock
\showISBNx{9798350328387}
\href{https://doi.org/10.1109/ISMAR59233.2023.00030}{doi:\nolinkurl{10.1109/ISMAR59233.2023.00030}}


\bibitem[Wang et~al\mbox{.}(2022)]%
        {wang_comprehensive_2022}
\bibfield{author}{\bibinfo{person}{Zhuo Wang}, \bibinfo{person}{Xiaoliang Bai}, \bibinfo{person}{Shusheng Zhang}, \bibinfo{person}{Mark Billinghurst}, \bibinfo{person}{Weiping He}, \bibinfo{person}{Peng Wang}, \bibinfo{person}{Weiqi Lan}, \bibinfo{person}{Haitao Min}, {and} \bibinfo{person}{Yu Chen}.} \bibinfo{year}{2022}\natexlab{}.
\newblock \showarticletitle{A comprehensive review of augmented reality-based instruction in manual assembly, training and repair}.
\newblock \bibinfo{journal}{\emph{Robotics and Computer-Integrated Manufacturing}}  \bibinfo{volume}{78} (\bibinfo{date}{Dec.} \bibinfo{year}{2022}), \bibinfo{pages}{102407}.
\newblock
\showISSN{0736-5845}
\href{https://doi.org/10.1016/j.rcim.2022.102407}{doi:\nolinkurl{10.1016/j.rcim.2022.102407}}


\bibitem[Yan(2022)]%
        {yan_augmented_2022}
\bibfield{author}{\bibinfo{person}{Wei Yan}.} \bibinfo{year}{2022}\natexlab{}.
\newblock \showarticletitle{Augmented reality instructions for construction toys enabled by accurate model registration and realistic object/hand occlusions}.
\newblock \bibinfo{journal}{\emph{Virtual Real.}} \bibinfo{volume}{26}, \bibinfo{number}{2} (\bibinfo{date}{June} \bibinfo{year}{2022}), \bibinfo{pages}{465--478}.
\newblock
\showISSN{1359-4338}
\href{https://doi.org/10.1007/s10055-021-00582-7}{doi:\nolinkurl{10.1007/s10055-021-00582-7}}


\bibitem[Yang et~al\mbox{.}(2020)]%
        {yang_comparing_2020}
\bibfield{author}{\bibinfo{person}{Yumeng Yang}, \bibinfo{person}{Joyce Karreman}, {and} \bibinfo{person}{Menno De~Jong}.} \bibinfo{year}{2020}\natexlab{}.
\newblock \showarticletitle{Comparing the {Effects} of {Paper} and {Mobile} {Augmented} {Reality} {Instructions} to {Guide} {Assembly} {Tasks}}. In \bibinfo{booktitle}{\emph{2020 {IEEE} {International} {Professional} {Communication} {Conference} ({ProComm})}}. \bibinfo{publisher}{IEEE}, \bibinfo{address}{Kennesaw, GA, USA}, \bibinfo{pages}{96--104}.
\newblock
\showISBNx{978-1-72815-563-0}
\href{https://doi.org/10.1109/ProComm48883.2020.00021}{doi:\nolinkurl{10.1109/ProComm48883.2020.00021}}


\bibitem[Yuan et~al\mbox{.}(2024)]%
        {yuan_augmented_2024}
\bibfield{author}{\bibinfo{person}{Bowen Yuan}, \bibinfo{person}{{https://orcid.org/0000-0002-3012-4077}}, \bibinfo{person}{{View Profile}}, \bibinfo{person}{Hyunwoo Cho}, \bibinfo{person}{{https://orcid.org/0000-0002-8684-4989}}, \bibinfo{person}{{View Profile}}, \bibinfo{person}{Gun Lee}, \bibinfo{person}{{https://orcid.org/0000-0002-1644-6934}}, \bibinfo{person}{{View Profile}}, \bibinfo{person}{Mark Billinghurst}, \bibinfo{person}{{https://orcid.org/0000-0003-4172-6759}}, {and} \bibinfo{person}{{View Profile}}.} \bibinfo{year}{2024}\natexlab{}.
\newblock \showarticletitle{Augmented {Reality} {Spatial} {Guidance} {Cues} for {Flexible} {Multi}-{Objective} {Task} {Environments}}.
\newblock In \bibinfo{booktitle}{\emph{Proceedings of the 19th {ACM} {SIGGRAPH} {International} {Conference} on {Virtual}-{Reality} {Continuum} and its {Applications} in {Industry}}}. \bibinfo{pages}{1--9}.
\newblock
\showISBNx{9798400713484}
\href{https://doi.org/10.1145/3703619.3706047}{doi:\nolinkurl{10.1145/3703619.3706047}}


\bibitem[Zhang et~al\mbox{.}(2025)]%
        {zhang_learning-based_2025}
\bibfield{author}{\bibinfo{person}{Xingjian Zhang}, \bibinfo{person}{Yutong Duan}, {and} \bibinfo{person}{Zaishu Chen}.} \bibinfo{year}{2025}\natexlab{}.
\newblock \bibinfo{title}{Learning-based {Stage} {Verification} {System} in {Manual} {Assembly} {Scenarios}}.
\newblock
\href{https://doi.org/10.48550/arXiv.2507.17304}{doi:\nolinkurl{10.48550/arXiv.2507.17304}}
\newblock
\shownote{arXiv:2507.17304 [cs]}.


\end{thebibliography}

\end{document}